\documentclass[11pt,twoside]{article}
\usepackage{FUSE2004}
\usepackage{natbib}

\usepackage{epsf}
\usepackage{psfig}
\usepackage{lscape}

\begin{document}
\title{An Unbiased Far Ultraviolet Survey of Magellanic Cloud Supernova Remnants}
\author{Parviz Ghavamian\altaffilmark{1}, William P. Blair\altaffilmark{1}, Ravi Sankrit\altaffilmark{1}, 
Charles Danforth\altaffilmark{2} \& Kenneth Sembach\altaffilmark{3}}
\affil{
$^1$Department of Physics and Astronomoy, Johns Hopkins University, 3400 North Charles Street,
Baltimore, MD 21218-2686 \\
$^2$Center for Astrophysics and Space Astronomoy, University of Colorado, Campus Box 593, Boulder,
CO 80302 \\
$^3$Space Telescope Science Institute, 3700 San Martin Drive, Baltimore, MD 21218  }

\begin{abstract}

We have undertaken a FUSE survey of Magellanic Cloud supernova
remnants, looking primarily for O~VI and C~III emission lines.
Work in earlier cycles with FUSE indicates that
optical and/or X-ray characteristics of supernova remnants are not always good
predictors of the objects that will be bright and detectable in the UV.
The goal of our survey is to test this concept by obtaining spectra
of a random sample of Magellanic Cloud remnants with a broad range of
radio, optical, and X-ray properties. Previously observed objects and
remnants with known high extinction (or known high column densities) are
the only objects eliminated from consideration.  
To date, we have clearly detected O~VI emission from 12 of the
33 objects observed, with weak or marginal detections possible in 
a handful more. 

\end{abstract}

\section{Introduction}

In the study of emission line objects, the far ultraviolet (FUV) is typically viewed
as an extension of the optical.  Consequently, FUV observations tend to target objects
(or parts of objects) which exhibit strong optical line emission.
This criterion has been frequently used in the field of supernova remnant (SNR) research
to select suitable targets for ultraviolet observations.  There could be many 
Magellanic Cloud SNRs detectable
at a level useful for analysis in the FUV, but which have never been observed because of
selection biases.  
To avoid the selection bias of targeting bright optical/X-ray remnants, we are
currently performing a FUSE Guest Investigator survey of all known LMC and SMC SNRs,
excluding only those which have been previously observed or are known to exhibit high interstellar 
extinction.  Most of these targets have not previously been observed in the UV.

In our observations a standard exposure of 10 ks with the LWRS aperture is requested, allowing us
to detect O~VI emission up to 50 times fainter than observed in such bright SNRs as N49 (Blair et al. 2000).  The
target coordinates are at the radio or X-ray center of each SNR.  With this 
observing setup, even non-detections are of significant interest.
Most of the objects are from 1\arcmin\ to 5\arcmin\
in extent, which will fill the LWRS aperture (30\arcsec $\times$ 30\arcsec) with emission.  
This setup also lessens the impact of relative FUSE channel
misalignments and keeps the observations simple.  In addition, Doppler shifting
from the front and back sides of the SNR shells ($\sim\,$30$-$150 km s$^{-1}$) will
move their emission lines out from under the overlying absorption of
the host galaxy.  

\section{FUSE Spectra from the LMC Survey: Two Examples}

Our FUSE observations of the MC SNRs were performed 
between 2003 July 14 and 2004 July 14.  During this period, 32 LMC remnants 
and 1 SMC remnant were observed.  
O~VI emission was detected from {\bf 12} of the {\bf 33} objects observed.
The O~VI $\lambda$1032 profiles exhibit a wide range of shapes.  Some are broad, some narrow,
and some are double peaked.  Some SNRs exhibit both O~VI and C~III emission, while others
are only detected in O~VI.  Thus far the faintest detections in our survey are at flux
levels F(1032)\,$\approx$\,10$^{-14}$ ergs cm$^{-2}$ s$^{-1}$ and 
F(977)\,$\approx$\,1.4$\times$10$^{-14}$ ergs cm$^{-2}$ s$^{-1}$.  An interesting
result already emerging from this survey is that, as suspected, {\it FUV emission is
detected even in SNRs where the targeted region exhibits minimal optical and X-ray emission.}
\begin{figure}[!ht]
\begin{center}
\plottwo{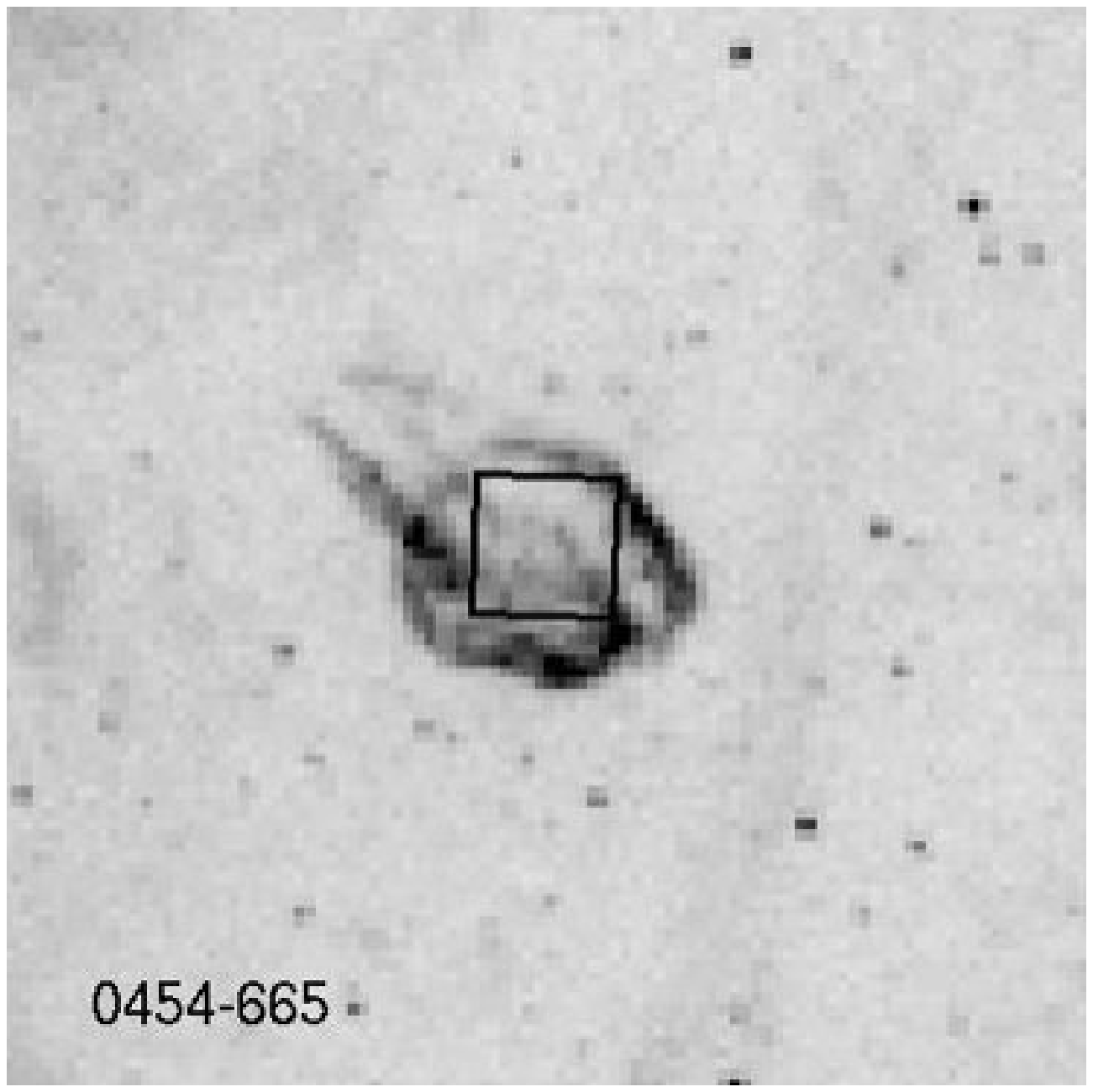}{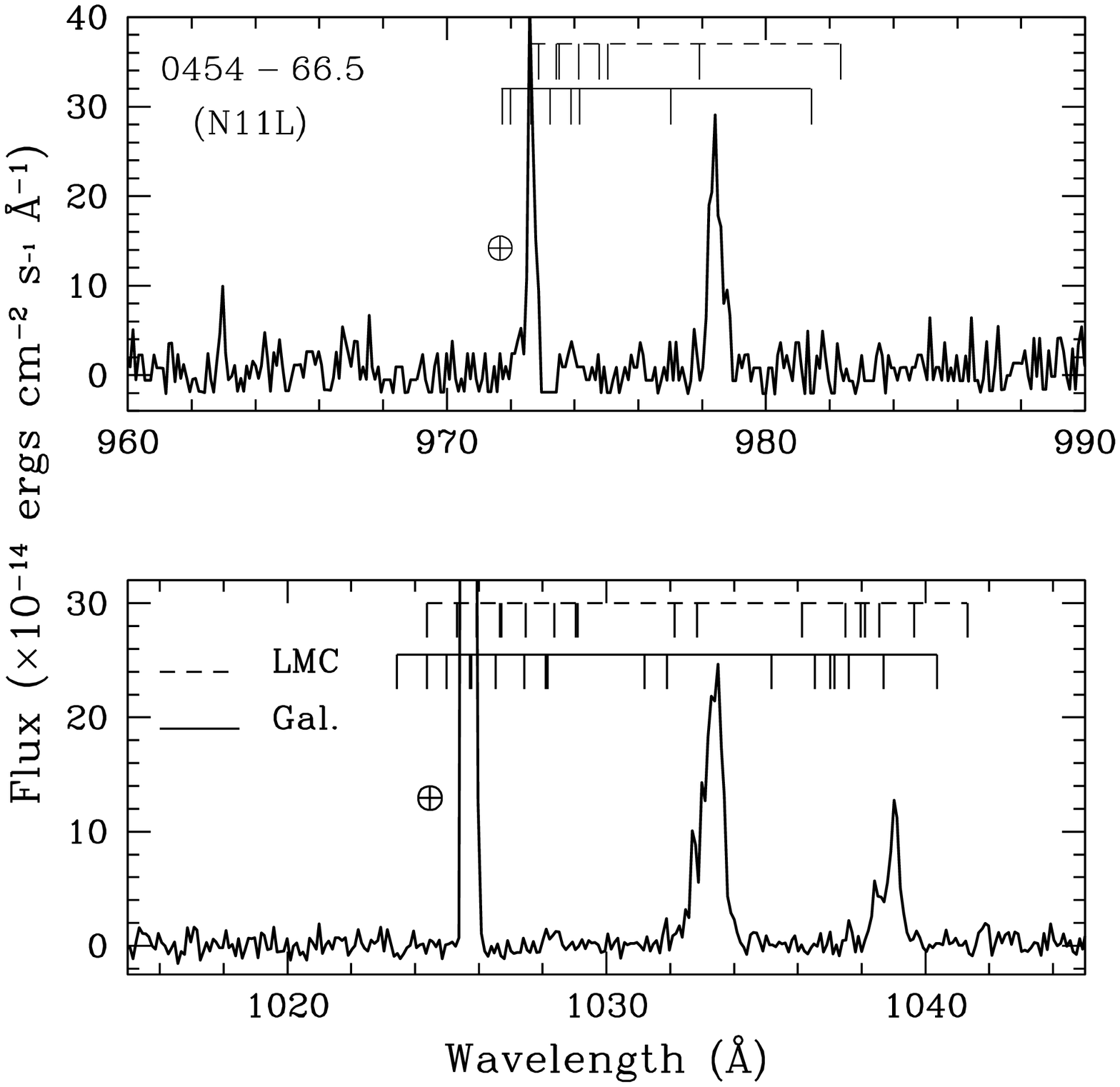}
\caption[]{Left: MCELS H$\alpha$ image of the LMC SNR 0454$-$66.5 (N11L).  Nominal placement of
the FUSE LWRS aperture is marked.  Right: The FUSE spectrum of 0454$-$66.5.  Bright
emission lines of shock-excited C~III $\lambda$977 and O~VI $\lambda\lambda$1032, 1038
are detected.  Positions of prominent Galactic/LMC absorption features are marked.  }
\label{snr0454}
\end{center}
\end{figure}

\begin{figure}[!ht]
\begin{center}
\plottwo{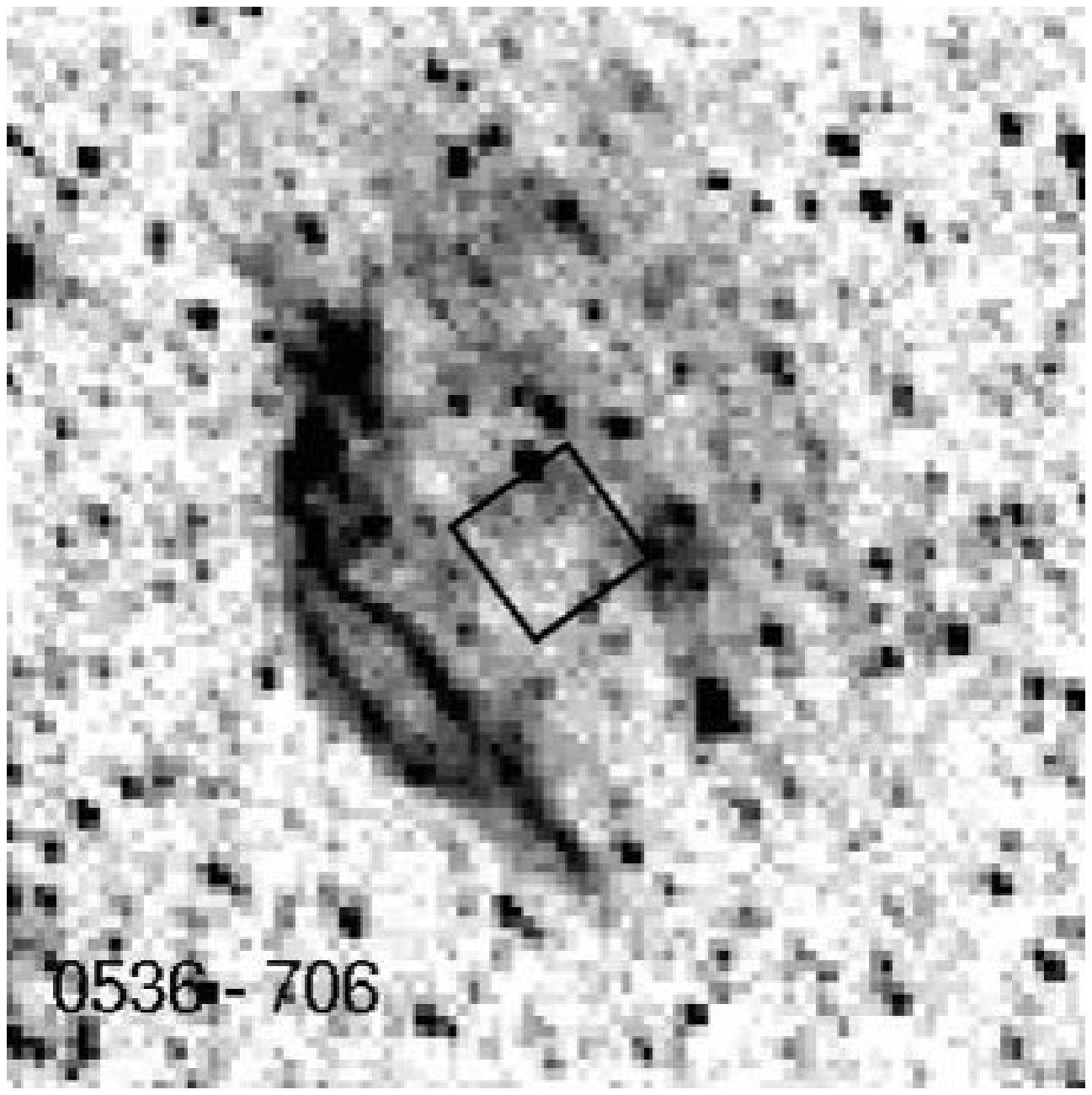}{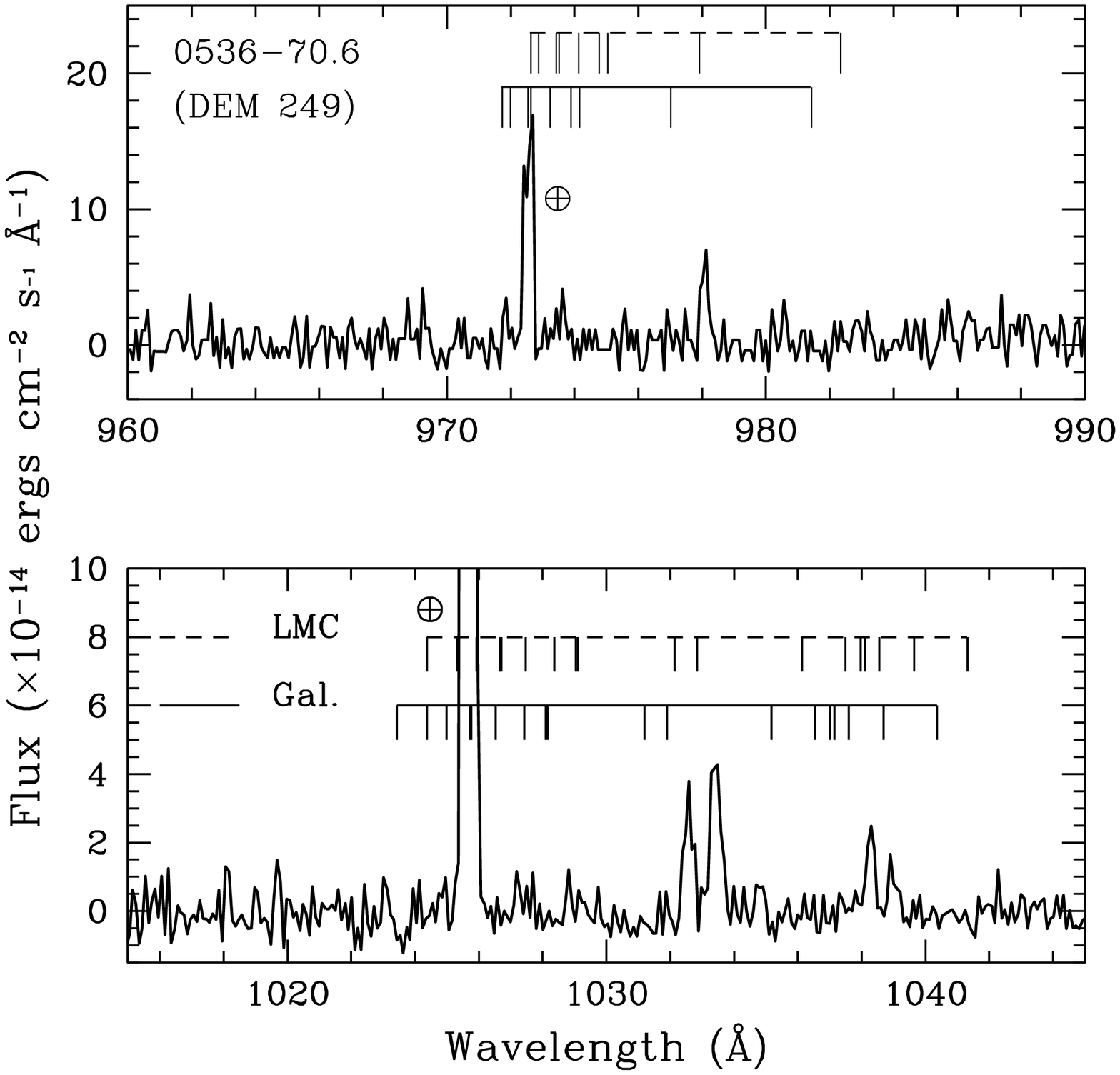}
\caption[]{Left: MCELS H$\alpha$ image of the LMC SNR 0536$-$70.6 (DEM 249).  Position of
the FUSE LWRS aperture is marked.  Right: The FUSE spectrum of 536$-$70.6.
Emission lines of shock-excited C~III $\lambda$977 and O~VI $\lambda\lambda$1032, 1038
are seen. The O~VI lines are double, suggesting that
emission is detected from the front and back sides of the SNR shell.  The C~III line is
redshifted from the LMC velocity, so it is likely that this component arises from the back side
of the shell.}
\label{snr0536}
\end{center}
\end{figure}

In Figs.~\ref{snr0454}\, and \ref{snr0536} we show H$\alpha$ images (from the Magellanic Cloud Emission Line
Survey (MCELS), Smith et al. 1999) and FUSE spectra of two "hits" from our sample LMC SNRs.
The diameters of these remnants 
differ significantly: 0454$-$66.5 (N11L; Henize 1956) and 0536$-$70.6 (DEM 249, Davies et al.
1976) are 1\arcmin\, (15 pc) and 2.\arcmin4 (36 pc) across, likely reflecting different ages.  The spectra
were produced from the orbital night data, with exposure 
times of 18.8 ks and 3.2 ks.   These two objects exhibit very different spectra: 0454$-$66.5
shows strong and broad O~VI and C~III emission (F(1032)$\approx\,$2$\times$10$^{-13}$ 
ergs cm$^{-2}$ s$^{-1}$, F(977)$\approx\,$1.5$\times$10$^{-13}$ ergs cm$^{-2}$ s$^{-1}$,
V$_{FWHM}$\,$\approx$\,400 km s$^{-1}$).  On the other hand, flux levels in 0536$-$70.6 are 
nearly ten times fainter, with each of the O~VI lines split into two components.
Here, the lines are approximately 150 km s$^{-1}$ wide and spaced approximately 230 km s$^{-1}$ apart (see Fig.~\ref{snr0536}).
\vspace{-0.3in}
\section{Summary and Future Directions} 
Thus far we have detected O~VI line emission from over 1/3 of the Magellanic Cloud
remnants observed in our survey.  This is an ongoing project, and will yield useful 
constraints on the kinematics of MC SNRs.
A broad assortment of data is available to support our FUSE observations:
Narrow band MCELS imaging of many LMC/SMC fields of interest are in hand,
while longslit echelle observations cutting across many
Magellanic Cloud H~II regions, bubbles, etc., are available to
provide an excellent base of general kinematic information on various
regions of potential interest.  X-ray observations with {\it Chandra}, both in the
archive and in ongoing programs, will provide invaluable information on
shocked ejecta and blast wave emission in these SNRs.  In addition, upcoming
{\it Spitzer} Cycle 1 observations will provide supplemental information on
shocked dust in the LMC SNRs.   




\end{document}